\documentclass[cits]{PoS}
\usepackage{graphicx}
\usepackage{rotating}

\title{Powerful relativistic jets in narrow-line Seyfert 1 galaxies (review)}

\ShortTitle{Powerful relativistic jets in narrow-line Seyfert 1 galaxies (review)}

\author{\speaker{Luigi Foschini}\thanks{I would like to thank the organizers, and particularly S. Komossa, for having realized a pleasant and scientifically fruitful workshop, and for the financial support.}\\
        INAF -- Osservatorio Astronomico di Brera, Via E. Bianchi 46, 23807, Merate (LC), Italy\\
        E-mail: \email{luigi.foschini@brera.inaf.it}}

\abstract{The recent detection of high-energy gamma rays from Narrow-Line Seyfert 1 Galaxies has confirmed that also this type of active galactic nuclei can generate powerful relativistic jets. I outline the evolution of the knowledge in this research field and the implications on the unification of relativistic jets at all scales.}

\FullConference{Nuclei of Seyfert galaxies and QSOs - Central engine \& conditions of star formation\\
		 November 6-8, 2012\\
		 Max-Planck-Insitut f\"ur Radioastronomie (MPIfR), Bonn, Germany}

\begin{document}

\section{Early times}
In early-mid nineties, the unified model of active galactic nuclei (AGN) reached a reliable shape as the result of the efforts of many researchers (e.g. \cite{KEEL,BARTHEL,ANTONUCCI,URRY}). Narrow-Line Seyfert 1 Galaxies (NLS1), recognized as a peculiar class of AGN in the second half of eighties (\cite{POGGE,GOODRICH}, but see also \cite{GASKELL}), were initially not included in the unified model. In the Table~1 of Urry \& Padovani \cite{URRY} NLS1s are not present, but would have been likely to be placed in the radio-quiet branch, under the column Type 0 or between Type 1 and Type 0. Later works included NLS1s and placed them in a region of radio-quiet sources characterized by low mass of the central black hole, high accretion luminosity, and hosted by spiral galaxies (e.g. \cite{BOROSON}, Fig.~7). 

However, also a population of radio-loud NLS1s (RLNLS1s) emerged slowly, one at a time. The first RLNLS1 -- PKS 0558$-$504 ($z=0.137$) -- was discovered in 1986 by Remillard et al. \cite{REMILLARD1} during the identification of eight sources detected by \emph{HEAO 1}. It was not yet recognized as NLS1 (the seminal paper by Osterbrock \& Pogge \cite{POGGE} was published just the previous year), but they wrote: \emph{``One of the I Zw 1 types, the previously unidentified radio source PKS 0558-504, is a QSO with unusually narrow hydrogen lines for a high-luminosity object ($M_{v}=-25.1$).''} Later, again Remillard et al. \cite{REMILLARD2}, reported strong and rapidly variable X-ray emission ($+67$\% in 3 minutes) from the same source detected by \emph{Ginga}. The measured energies and time scales required relativistic beaming to be reasonably explained. A second candidate -- RGB J0044$+$193 ($z=0.181$) -- was found in 1999 while searching for X-ray selected BL Lac Objects in the \emph{ROSAT}-Green Bank survey (RGB, \cite{SIEBERT}), but it was later suggested that the radio detection could be spurious, thus restoring the radio-quiet classification \cite{MACCARONE}. More RLNLS1s were discovered in these years. Grupe et al. \cite{GRUPE} in 2000 reported about RX~J0134.2$-$4258 ($z=0.237$): \emph{ROSAT} observations indicated spectral variability (fainter when softer), with a variable hard component, confirmed also by \emph{ASCA}. They proposed three possible explanations: warm absorber, corona, or relativistic jet. The third case followed soon: PKS~2004$-$447 ($z=0.24$) was discovered by Oshlack et al. \cite{OSHLACK} (see also \cite{GALLO}). About one dozen of RLNLS1s were found in a survey done by Zhou \& Wang \cite{ZHOU1}. 

The advent of the {\it Sloan Digital Sky Survey} (SDSS) in early 2000s determined a major change in this research field. The public availability of thousands of optical spectra allowed the cross-correlation with radio catalogues to search for more candidates. Particularly, it is worth mentioning H.~Y. Zhou, who gave many important contributions by discovering, with his colleagues, several RLNLS1s \cite{ZHOU2,ZHOU3,ZHOU4,YUAN}. Specifically, they reported about the radio properties of three sources -- SDSS~J094857.3$+$002225 ($z=0.585$) \cite{ZHOU2}, SDSS J084957.97$+$510829.0 ($z=0.584$) \cite{ZHOU3}, and 1H~0323$+$342 ($z=0.061$) \cite{ZHOU4} -- that were very similar to blazars (flat or inverted radio spectrum, high brightness temperature), suggesting the presence of a relativistic jet viewed at small angles. Interestingly, all these three sources were later detected at high-energy $\gamma$ rays \cite{LAT4}. These works triggered deeper studies at radio frequencies of some RLNLS1s \cite{DOI1,DOI2}, where the typical radio properties of relativistic jets were confirmed, and, in addition, the radio morphology resulted to be very compact on parsec scale.

A first attempt to search for emission at TeV energies from RLNLS1s was done by Falcone et al. \cite{FALCONE} in 2004 by using the ground-based \emph{Whipple} telescope. However, the aim of that study was to detect at $\gamma$ rays some candidates of the elusive population of high-frequency peaked flat-spectrum radio quasars (HFSRQs) postulated by Padovani et al. \cite{PADOVANI1,PADOVANI2} to challenge the so-called ``blazar sequence'' \cite{FOSSATI,GHISELLINI2}. According to the latter, blazars follow a sequence linking the frequency of the synchrotron peak to the jet power: the greater the power, the lower the synchrotron peak frequency, and vice versa. Therefore, to find a high-power blazar with a high synchrotron peak frequency (in the UV/X-rays) would have been a major break in the sequence. Among the candidates in the list of Falcone there were two RLNLS1s -- 1H~0323$+$342 and SDSS~J162901.30$+$400759.9 -- but not recognized as such. It was likely a misinterpretation of the strong soft X-ray emission of NLS1s, due to the accretion disk, which instead was considered as an indication of a synchrotron peak at X-rays\footnote{The HFSRQs were at last found in 2012 by Padovani et al. \cite{PADOVANI3}. See, however, a different interpretation by Ghisellini et al. \cite{GHISELLINI1} and Foschini \cite{FOSCHINI9}.}. Anyway, the search of TeV emission was negative, but intriguing. In the words by Falcone et al. \cite{FALCONE}: \emph{``No significant emission has been detected from any of the candidate sources in this initial survey. There was marginal evidence of a rate increase observed in the B2~0321$+$33 {\rm [alias 1H~0323$+$342]} light curve, but the statistical significance of this increase is $2.5\sigma$ (post-trial significance), which could be accounted for by a statistical fluctuation''}.

In mid-2000s, two more surveys were done by Whalen et al. \cite{WHALEN} and Komossa et al. \cite{KOMOSSA1}, who did also a specific paper on one source of her list \cite{KOMOSSA2}. The former survey was done by using the {\it FIRST Bright Quasar Survey} (FBQS) and the authors conclude that \emph{``except for their radio properties, radio-selected NLS1 galaxies do not exhibit significant differences from traditional NLS1 galaxies. Our results are also in agreement with previous studies suggesting that NLS1 galaxies have small black hole masses that are accreting very close to the Eddington rate''} \cite{WHALEN}. Instead, Komossa et al. studied a more heterogeneous and small sample and concluded that \emph{``while properties of most sources (with two to three exceptions) generally do not favor relativistic beaming, the combination of accretion mode and spin may explain the observations''} \cite{KOMOSSA1}. In 2008, there was the SDSS sample of 23 RLNLS1s by Yuan et al. \cite{YUAN}: \emph{``Intrinsically, some of them have relatively low radio power and would have been classified as radio-intermediate AGNs. The black hole masses are estimated to be within $10^{6}-10^{8}M_{\odot}$, and the Eddington ratios close to unity, as in normal NLS1 AGNs. The results imply that radio-loud AGNs may be powered by black holes of moderate masses ($\sim 10^{6}-10^{7}M_{\odot}$) accreting at high rates''}. 

\begin{figure}[!ht]
\begin{center}
\includegraphics[angle=270,scale=0.28,clip,trim=20 50 30 50]{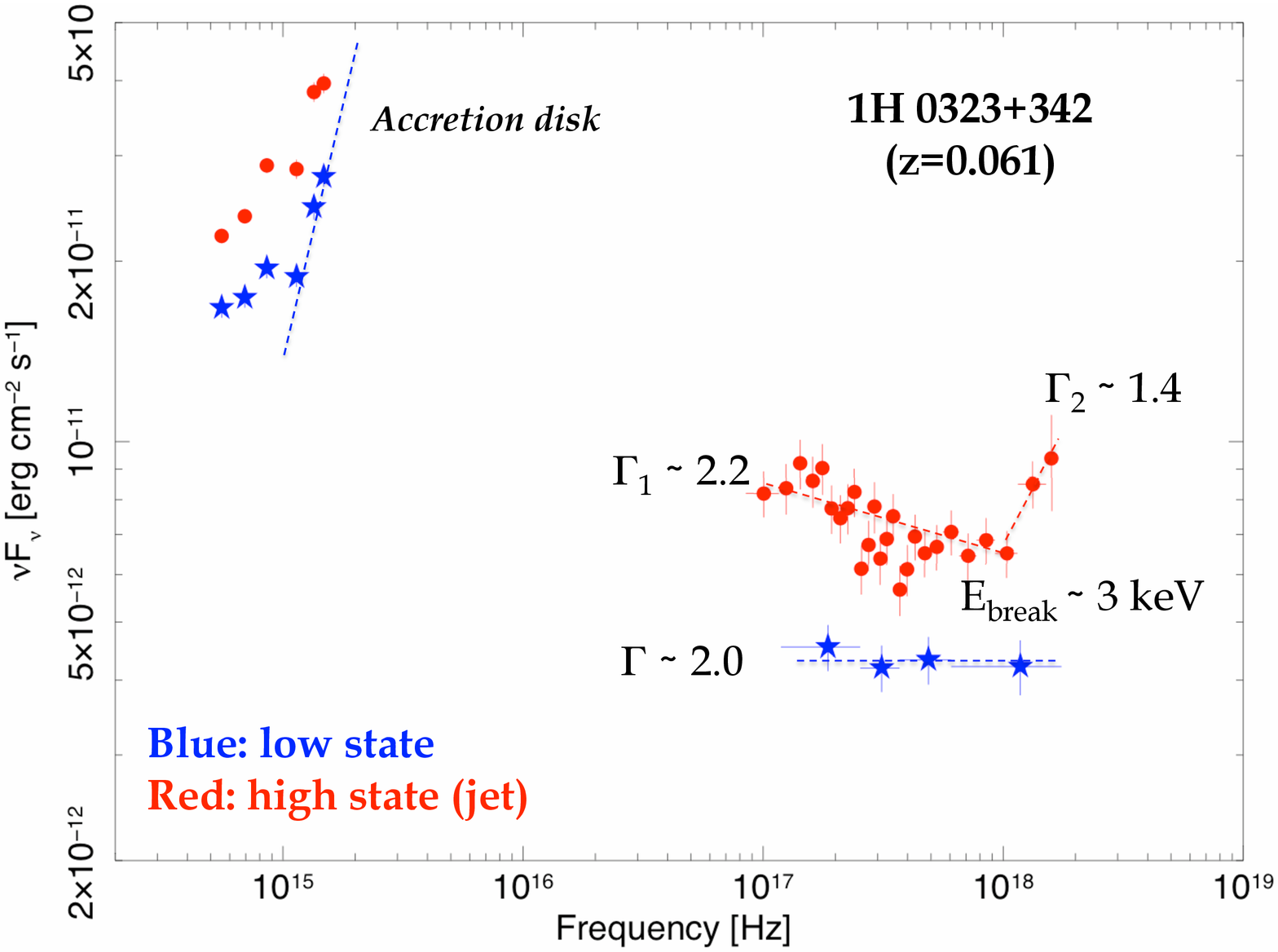}
\includegraphics[angle=270,scale=0.28,clip,trim=30 50 30 50]{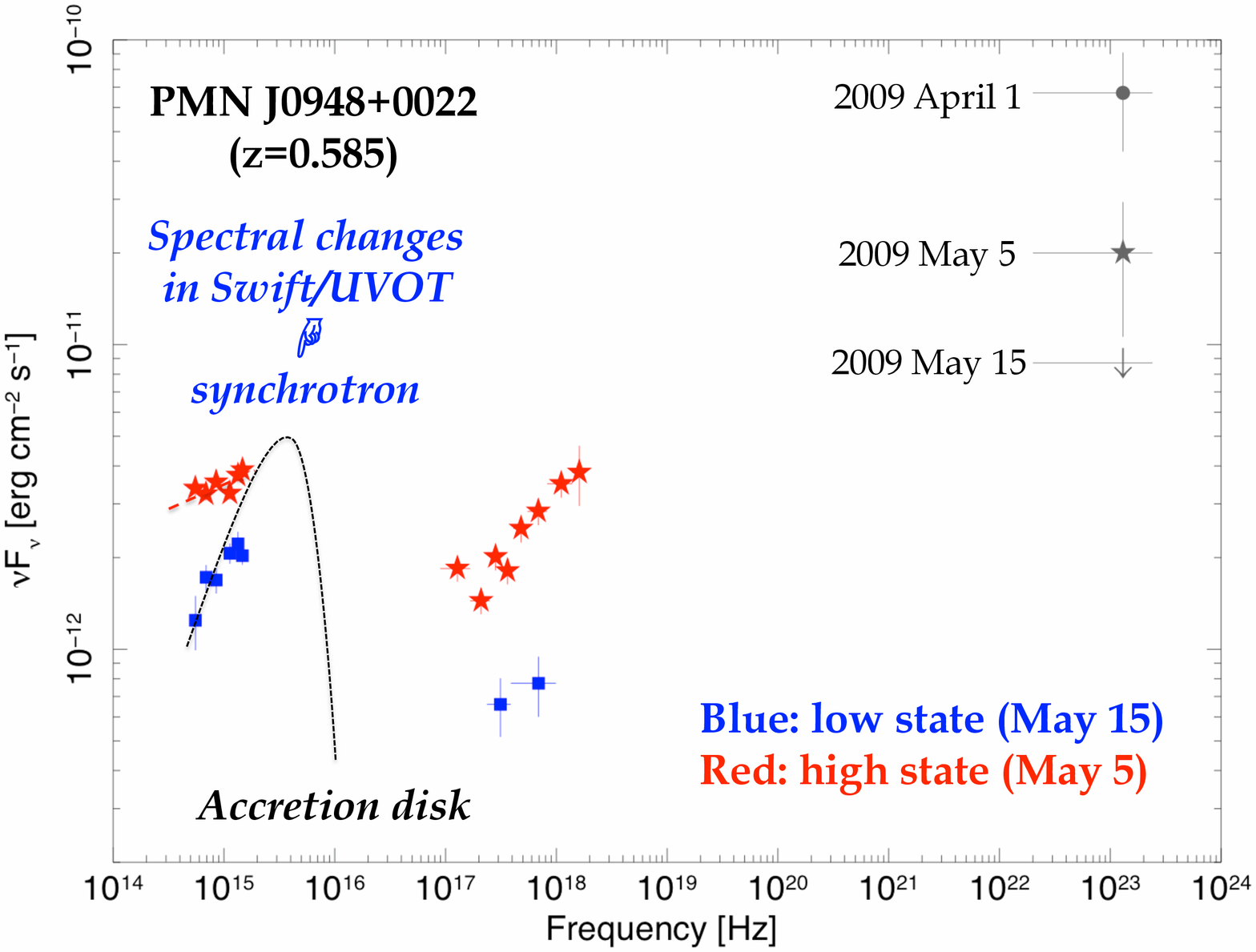}
\caption{(\emph{left panel}) \emph{Swift} (XRT and UVOT) observations of the RLNLS1 1H~0323$+$342 in two different states. When the jet is not active (blue points, ObsID 00036533007, 21 Dec 2007), the source displays low optical-to-X-ray flux, with a rather flat X-ray photon index. As the jet increases its activity (red points, ObsID 00035372001, 6 Jul 2006), there is a greater optical-to-X-ray flux and the emergence of a hard tail ($\Gamma \sim 1.4$). See also \cite{FOSCHINI1}. (\emph{right panel}) \emph{Swift} (XRT and UVOT) and \emph{Fermi}/LAT observations of the RLNLS1 PMN~J0948$+$0022 in two different states during the 2009 MW campaign \cite{LAT3}. The source was active during 2009 April, with a peak on the day 1. Then, in early May, the activity dropped at all the wavelengths in a coordinated way. See \cite{LAT3} and the text for more details.}
\label{fig:SWIFT}
\end{center}
\end{figure}

On the basis of the radio morphology and spectra, Komossa et al. \cite{KOMOSSA1} also suggested that some RLNLS1s of their sample could be similar to Compact Steep Spectrum (CSS) radio sources. Gallo et al. \cite{GALLO} found that one clear example of RLNLS1/CSS was PKS~2004$-$447. Although Komossa et al. \cite{KOMOSSA1} cited this source, they did not include in their sample on the basis of a weak FeII bump, which in turn would suggest a classification of this AGN as narrow-line radio galaxy or type II AGN. However, Gallo et al. \cite{GALLO} noted that there is no formal threshold for the intensity of Fe II, thus holding the classification as RLNLS1.

Meanwhile, another important player introduced itself in this game: the \emph{Swift} satellite, launched on November 2004. Again, the policy of use played an important role. The possibility for almost everyone to have even a little snapshot and the immediate public availability of data, made it possible to perform many researches outside the mainstream, as it was the case of RLNLS1s. In the second half of 2000s, the {\it Swift} public archive was sufficiently rich of observations on RLNLS1 to perform an early little survey \cite{FOSCHINI1}. Some interesting results were found, specifically about 1H~0323$+$342 (Fig.~\ref{fig:SWIFT}, \emph{left panel}): the source had a strong optical-to-X-ray emission from the accretion disk, but sometimes the jet emerged, resulting in the appearance of a hard tail. This behaviour was confirmed also by the spectral variability at hard X-rays. The source was observed by \emph{INTEGRAL} in 2004 with a low flux and soft spectrum ($F_{20-40\ \rm keV}= 2.5\pm0.5$~mCrab; $F_{40-100\ \rm keV} < 2.6$~mCrab), while the integration of the {\it Swift}/BAT on axis data between 2006-2008 (exposure $\sim 53$~ks) resulted in a high flux and hard spectrum ($F_{20-40\ \rm keV}< 20$~mCrab; $F_{40-100\ \rm keV} = 16\pm2$~mCrab) \cite{FOSCHINI1}. 

Also the hard X-ray detection was affected by a misinterpretation. Indeed, 1H~0323$+$342 was already present in a few papers on hard X-ray sources published in 2007, before of \cite{FOSCHINI1}: two independent catalogs of the IBIS imager onboard the \emph{INTEGRAL} satellite \cite{BIRD,KRIVONOS} and one optical-to-X-ray follow-up with \emph{Swift} of a sample of 34 hard X-ray AGN \cite{MALIZIA1}. However, in all these papers, 1H~0323$+$342 was not recognized as RLNLS1, but classified as a normal Seyfert 1. In a subsequent paper reporting the \emph{``first high-energy observations of narrow-line Seyfert 1s''} -- where ``high-energy'' here means in the range 17-100~keV -- Malizia et al. \cite{MALIZIA2} studied two-three candidate radio-quiet NLS1s discovered by \emph{INTEGRAL} and \emph{Swift}/BAT, and with no reference to any rethinking about 1H~0323$+$342. The latter was included in the \emph{INTEGRAL} sample only in 2011 by Panessa et al. \cite{PANESSA}.

\section{The Fermi breakthrough}
A real change in the perception of RLNLS1s came in 2008 with the launch of the \emph{Fermi Gamma-ray Space Telescope} (hereafter \emph{Fermi}), but also in this case the emerging of RLNLS1s as a new class of $\gamma$-ray emitting AGN was not straightforward. Indeed, the first RLNLS1 to be detected at MeV-GeV energies -- SDSS~J094857.3$+$002225 (alias PMN~J0948$+$0022) \cite{LAT2} -- was also present among the list of bright sources detected after the first three months of \emph{Fermi} activity \cite{LAT1}. However, it was still identified as a FSRQ, although the text refers to another specific paper in preparation: \emph{``The source PMN~J0948$+$0022, associated with 0FGL~J0948.3+0019, has a flat radio spectrum but shows an optical spectrum with only narrow emission lines, leading to an `uncertain' type classification in Roma-BZCat. A detailed analysis of this source is presented in Abdo et al. (2009a)''} [the latter being the reference \cite{LAT2} in the present work]. It was indeed in \cite{LAT2} that it was reported the first high-energy $\gamma$-ray detection of a RLNLS1 in an explicit and ``conscious'' way. The discovery was soon followed by a complementary paper containing additional information to improve the identification \cite{LAT5}. 

Obviously, these early works linked the $\gamma$-ray source with the RLNLS1 on a statistical basis. Therefore, a multiwavelength (MW) campaign was activated to study in detail the electromagnetic emission of PMN~J0948$+$0022 from radio to $\gamma$ rays \cite{LAT3}. We had sufficient luck to catch coordinated variability at different frequencies. PMN~J0948$+$0022 displayed a moderate $\gamma$-ray activity on 2009 April, with fluxes of the order of a few $\times 10^{-7}$~ph~cm$^{-2}$~s$^{-1}$ ($0.1<E<100$~GeV). In early May there was a drop of the $\gamma$-ray flux followed by a decrease of the optical-to-X-ray flux as measured by \emph{Swift}, together with a spectral change in the optical/UV spectrum (Fig.~\ref{fig:SWIFT}, \emph{right panel}). This was interpreted as a decrease of the jet emission (synchrotron), which left the optical/UV wave band dominated by the thermal emission from the accretion disk. This point of view was confirmed by the fact that a few weeks later, there was the peak of the radio emission as recorded from OVRO, Mets\"ahovi, and Effelsberg ground-based radio telescopes \cite{LAT3}, as expected from the classical theory of relativistic jets \cite{BLANDFORD}. Indeed, the blob of plasma has to be compact to be optically thin to $\gamma$ rays, but in this case the lower part of the electromagnetic spectrum (radio frequencies) is self-absorbed. As the blob moves outward, it expands itself, thus becoming optically thick to $\gamma$ rays, but thin to radio emission. Moreover, additional support to this interpretation came from the detection of optical ($V$) polarization at $\sim 19$\% from the Kanata telescope between the end of March and the beginning of April 2009, when the source was moderately active at $\gamma$ rays \cite{IKEJIRI}. All these information concurred to establish a well-grounded association of the \emph{Fermi} detected $\gamma$-ray source with the RLNLS1 PMN~J0948$+$0022. Other MW campaigns strengthened the association and the similarities of the jet with those of blazars \cite{FOSCHINI3,FOSCHINI4}. Other optical-infrared observations confirmed violent intranight variability, as expected from AGN with relativistic jets viewed at small angles \cite{LIU,JIANG,PALIYA}.

After one year of \emph{Fermi} operations, the number of $\gamma$-ray detected RLNLS1s increased to four \cite{LAT5}, with the addition of 1H~0323$+$342 (cf \cite{ZHOU4,FOSCHINI1,PANESSA}), PKS~2004$-$447 (cf \cite{OSHLACK,GALLO,FOSCHINI1}), and PKS~1502$+$036 (cf \cite{YUAN}). RLNLS1s were then explicitly indicated \emph{``as a new class of gamma-ray active galactic nuclei''}, in addition to blazars and radio galaxies \cite{LAT5}. One more RLNLS1, SBS~0846$+$513 (cf. \cite{ZHOU3}) was later detected because of an outburst (\cite{FOSCHINI2,LAT6,DAMMANDO}), indicating that the probability of detection at $\gamma$ rays is still biased by the sensitivity of the instrument. The above cited sources have all been detected at $\gamma$ rays with high significance ($TS>25$, see \cite{MATTOX} for the definition of TS, or $\gtrsim 5\sigma$). However, it is worth noting that there are also several detections at low significance ($9<TS<25$ or $3<\sigma<5$), so that there is presently one dozen of RLNLS1s detected at MeV-GeV energies \cite{FOSCHINI2} (\footnote{See the updated list at the web page {\tt http://tinyurl.com/gnls1s}}). The number is expected to increase, depending on the activity of the sources. The ability of {\it Fermi} to scan all the sky every three hours guarantees a continuous monitoring.

\section{The parent population}
One dozen of RLNLS1s with the jet viewed at small angles (five, in the most conservative hypothesis) implies $\sim 2\Gamma^2$ times parent sources, i.e. with the jet viewed at large angles. $\Gamma$, the bulk Lorentz factor of the jet, is generally $\sim 10$, which in turn means at least $\sim 10^{3}$ parent sources. The problems is that it is quite difficult to find them. 

\begin{figure}[!ht]
\begin{center}
\includegraphics[angle=270,scale=0.5]{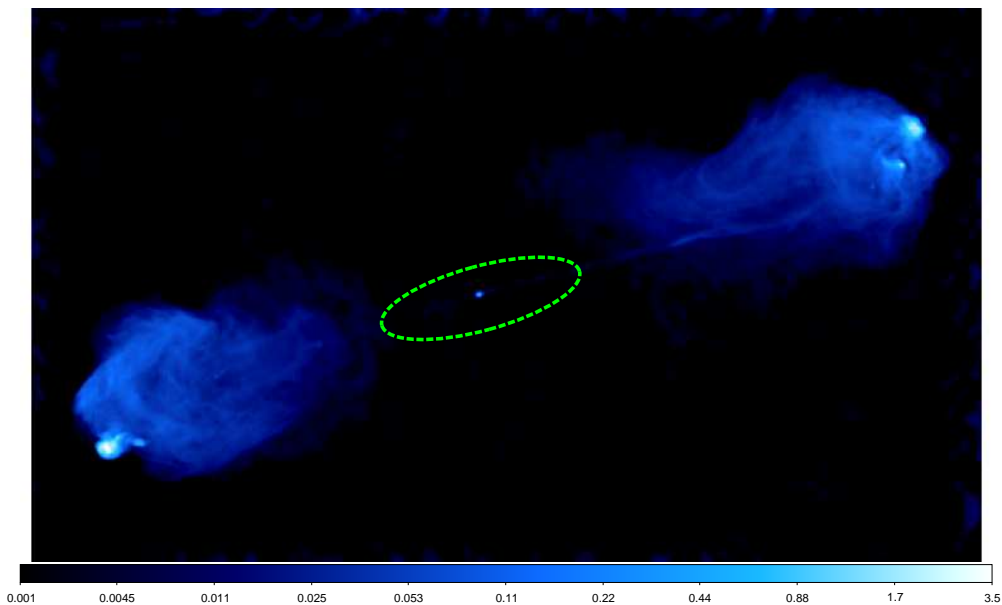}
\caption{Cygnus A as observed by the Very Large Array (VLA) at 6~cm \cite{PERLEY}. Personal elaboration of the original FITS file. See text for details.}
\label{fig:CYGNUSA}
\end{center}
\end{figure}

Basically, there are three hypotheses:

\begin{enumerate}
\item The first one -- rather obvious -- is to search simply for RLNLS1s with jet viewed at large angles. This could be not so easy, because early studies indicated a very compact structure for beamed NLS1s \cite{DOI1,DOI2,GU,GIROLETTI1}. Nevertheless, some interesting cases are emerging, like PKS~0558$-$504, where Gliozzi et al. \cite{GLIOZZI} found a bipolar radio jet 46~kpc-long (projected) likely viewed at $30^{\circ}-45^{\circ}$. Three more candidates have been found by Doi et al. \cite{DOI3}.

\item The second hypothesis is based on the fact that RLNLS1s are extremely compact at radio frequencies. If there is no extended emission, then it is possible that these sources are radio quiet when observed at large angles. To better understand this hypothesis, have a look at one well-known radio image of the radio galaxy Cygnus~A (Fig.~\ref{fig:CYGNUSA}). Most of the radio emission comes from the hot spots and the core, while the jet is almost invisible, an indication of its high efficiency. Now think to a similar source, but without the hot spots. NLS1s are thought to be young AGN -- cf \cite{GRUPE2,MATHUR1,MATHUR2,KOMOSSA1} -- and therefore one could expect that the spots have not yet developed. Roughly, the source is now limited to what is inside the dashed ellipse in Fig.~\ref{fig:CYGNUSA}. The source observed pole-on has a compact morphology with a high brightness temperature core. When it is observed edge-on, the radio emission - no more enhanced by the special relativity effects - becomes much fainter and such a source might seem radio quiet. In this case, the parent population could be that of the radio-quiet NLS1s. Radio quiet, but not radio silent, as proved by \cite{ULVESTAD}. Indeed, some jet-like radio structures have been found also in radio-quiet NLS1s \cite{GIROLETTI2,MUNDELL,DOI4}.

\item The third and last option is based on the hypothesis that the broad-line region has a disk-like shape \cite{DECARLI}. This means that when the source is observed pole-on, there is no component of motion directed toward the observer and, hence, no Doppler broadening. When observed edge-on, the Doppler broadening is present and therefore the line profile is broad. Therefore, in this case, it is necessary to search among the usual radio galaxies, but hosted by disk/spiral galaxies, as NLS1s generally have that type of host\footnote{Blazars and radio galaxies are instead hosted by elliptical galaxies.}. Some examples have been found, e.g. \cite{BRUNTHALER,EMONTS,MCHARDY,PERLMAN,KEEL2,MUNDELL,MORGANTI}. It is worth noting that a systematic study on the morphology of the host galaxy of a flux-limited sample of radio galaxies (2~Jy) resulted in the evidence that 12\% of the sources are hosted by disk galaxies \cite{INSKIP}.
\end{enumerate}

The parent population presently remains an open question. It is not yet clear if one of the above three will win or if it will result a mixture including a bit of all the three or, even, in a yet-to-be-made fourth hypothesis. 

\section{The role of RLNLS1s in the unification of jets at all scales}
The discovery of powerful relativistic jets in RLNLS1s is not ``simply'' the addition of one more class of $\gamma$-ray AGN. It has deep implications in the unification of jets at all scales. RLNLS1s are different from blazars in many aspects, but the jets seem to be almost the same, as proved by the MW campaigns \cite{LAT3,FOSCHINI3,FOSCHINI4}. Indeed, some researchers (e.g. \cite{DAMMANDO}) suggested that RLNLS1s might be  blazars in a early stage of their life. While there is agreement on the physical properties of the jets in blazars and RLNLS1s, to name all these sources simply as blazars could be strongly misleading. The reason supporting the inclusion of RLNLS1 in the blazar name is that some researchers refer the term simply to the jet emission boosted by the special relativity (somehow derived from the verb ``to blaze''), disregarding any information on the AGN, environment, and host. Ulrich et al. \cite{ULRICH} wrote: \emph{``In radio-loud AGN seen at small angles to the axis of the jet, the highly nonthermal radiation produced in the jet is strongly amplified by relativistic beaming and dominates the observed continuum. In these sources, called blazars, variability is the most violent and affects the whole electromagnetic range from the radio to the gamma-ray band.''}. Burbidge \& Hewitt \cite{BURBIDGE} wrote that \emph{``at the dinner at the end of that meeting {\rm [the well-known Pittsburgh conference on BL Lac Objects held in 1978]}, Spiegel coined the term `blazar' a pictorial term which he proposed be applied to rapidly variable objects some of which, but not all, can also be classified as BL Lac objects.''} The problem is not the term itself: since it is now evident that the powerful jets in RLNLS1s are the same of blazars, it might be useful to speak about all these sources dominated by the jet emission as blazars, given the above cited definitions (\footnote{I myself coauthored in 2009 a paper titled: \emph{``Blazar nuclei in radio-loud narrow-line Seyfert 1?''} \cite{FOSCHINI1}.}).

The problem is that historically blazars refer to BL Lac Objects and FSRQs. According to \emph{Wikipedia}, the Spiegel's term was the result of the contraction of the words BL Lac objects and Optically Violent Variable (OVV) quasars(\footnote{{\tt http://en.wikipedia.org/wiki/Blazar}}). However, BL Lac Objects and FSRQs are AGN different from RLNLS1s in several ways. Therefore, the simple inclusion of RLNLS1s in the blazar semantic field results in missing some important new information specifically linked to these peculiar sources, which can be particularly useful for the unification of jets at all scales. For example, one is the confirmation of what already told Roger Blandford at the Pittburgh 1978 conference on BL Lac Objects: \emph{``As the continuum {\rm [jet]} emission is proposed to originate in the central 10 pc, I don't think the nature of the surrounding object is particularly relevant to the model''} \cite{BLANDFORD3}. The fact that powerful relativistic jets develop also in the environment of RLNLS1s, which is different from that of blazars, is one more arrow for the Blandford's bow. Another important information is the break down of the mass requirement of the central accreting object to develop a jet in AGN (e.g. \cite{LAOR}): since the masses of the central black hole of RLNLS1s are smaller than those of blazars, this requirement is no more valid, making thus possible to perform the unification (such mass requirement was not present on Galactic scale). 

Another change in the common knowledge is required to unify efficiently the jets at all scales and refers to the analogy with compact objects on Galactic scale. Traditionally, NLS1s have been considered as the large scale version of stellar mass black holes in soft/high state, because of the high-accretion rate (e.g. \cite{GLIOZZI}). However, what it matters is the mass not the accretion: when displaying the jet power as a function either of the mass of the central compact object or the disk luminosity, two branches resulted (see Fig.~3 in \cite{FOSCHINI5}). One dependent on the mass (RPD, radiation-pressure dominated regime) and the other dependent on the accretion (GPD, gas-pressure dominated regime), which in turn are in agreement with the expectations from the theory of jets \cite{BLANDFORD2,MODERSKI,GHOSH}. RLNLS1s are placed in the RPD branch. Another way to see this effect is displayed in Fig.~2 of \cite{FOSCHINI7}: it is evident that the RLNLS1s (low-mass AGN) make a branch similar to that of neutron stars on Galactic scale (low-mass binaries). It is also worth stressing that without powerful jets in RLNLS1s, the low-mass AGN branch would be missing, thus making impossible a unification of jets at all scales, despite the interesting attempts made in the past. Instead, now with RLNLS1, one can rescale the jet power according to the mass of the compact object ($M^{1.4}$ according to \cite{HEINZ}), making it possible to merge the Galactic and extragalactic jets \cite{FOSCHINI6,FOSCHINI7}. It remains to understand a slight dependence on the disk luminosity (cf. \cite{FOSCHINI6}), which in turn depends on our understanding of the disk structure and its efficiency in the conversion of the gravitational potential energy into radiation.

Another open question is the role of the spin of the compact object, although rather than the spin alone, it is necessary to study also how to measure the angular speed of the magnetic field lines (cf. \cite{BLANDFORD2}). As it will be possible to measure both angular velocities, perhaps it will be possible to understand the division between AGN with or without relativistic jets (e.g. \cite{FOSCHINI8}).

\section{Conclusions}
The class of RLNLS1 is an important piece in our understanding of relativistic jets at all scales. Although each work has given an important contribution to the present mosaic, the real breakthrough occurred in 2008 with the detection of high-energy $\gamma$ rays with \emph{Fermi}/LAT and the subsequent MW campaigns. This definitely proved the presence of powerful relativistic jets in this type of AGN and allowed to set an important step toward the unification of relativistic jets.

\end{document}